\documentstyle[psfig,conf_iap,]{article}
\begin{document}
\heading{%
%
Millions of Single Cloud Weak Mg II Systems
%
} 
\par\medskip\noindent
\author{Chris Churchill$^{1}$, Jane Rigby$^{2}$, Jane Charlton$^{1}$}
\address{Penn State, University Park, PA, 16802, USA}
\address{Steward Observatory, Tucson, AZ, 85721, USA}

\begin{abstract}
We report on a population of absorption systems selected by the
presence of very weak Mg~II doublets.  A sub--population of these
systems are iron enriched and have near solar metallicities.  This
would indicated advanced stages (i.e.\ few Gyr) of in situ star
formation within the absorbing structures.  From photoionization
modeling, we infer low ionization fractions of $f({\rm H~I/H_{tot}})
\simeq 0.01$, and gas densities of $\sim 0.1$~cm$^{-3}$.  Since the
maximum H~I column densities are $\sim 10^{17}$~cm$^{-2}$, the
inferred cloud sizes are $\sim 10$~pc.  From their redshift number
densities, this implies that their co--moving spatial density
outnumbers normal bright galaxies by a factor of a few million.
\end{abstract}

\section{What Good are Mg II Systems?}

The strong, resonant Mg~II $\lambda \lambda 2796, 2803$ doublet
provides a sensitive tracer of gas in the universe, sampling
environments over five decades of neutral hydrogen.  Mg~II allows the
cosmological tracking of damped Ly~$\alpha$ absorbers (DLAs, $N({\rm
H~I}) > 10^{20.3}$~cm$^{-3}$), Lyman limit systems (LLS, $N({\rm H~I})
> 10^{17.3}$~cm$^{-3}$), and so--called sub--Lyman limit systems
(sub--LLS, $N({\rm H~I}) < 10^{16.8}$~cm$^{-3}$).  Mg~II also traces
the presence of early--epoch star formation, since magnesium is a product of
Type~II supernovae (SNe), enriching surrounding absorbing gas clouds
on time scale of a few Myr.

Often, Fe~II and C~IV transitions are present in Mg~II ``selected''
systems; their strengths relative to Mg II and each other vary.
The presence of iron in some systems suggests more advanced star
formation processes, such as Type~Ia SNe that enrich their
surroundings on time scales of a few Gyr.  Strong C~IV in some systems
suggests density structure in the clouds, since Fe~II and C~IV cannot
simultaneously survive photoionization by the similar radiation
fields [see Churchill, et~al.\ (1999) and Rigby,
Charlton, \& Churchill (2001)].

\section{Observational Properties}

The population of Mg~II absorbers so--called ``weak systems'' are
isolated in redshift, have rest--frame equivalent widths less than
0.3~{\AA}, and exhibit single clouds with line widths unresolved at
HIRES/Keck resolution ($\Delta v = 6$~km/s).  In Figure~1, we present
examples of three single--cloud weak systems (SCWS) and one weak Mg~II
system with an additional very weak cloud on the blue wing ($z=1.229$).
The data been converted to rest--frame wavelengths.

For $W > 0.02$~{\AA}, the equivalent width distribution follows a
power law according to $n(W) = CW^{-1}$.  The redshift path density
outnumbers that of Lyman--limit systems by a factor of three; that is,
statistically 65\% of {\it all\/} Mg~II systems are optically thin in
neutral hydrogen  (Churchill et~al.\ 1999).  This has been verified
observationally by measuring the lack of a Lyman limit break in many
weak systems (Churchill et~al.\ 2000).

\begin{figure}[t]
\centerline{\vbox{
\psfig{figure=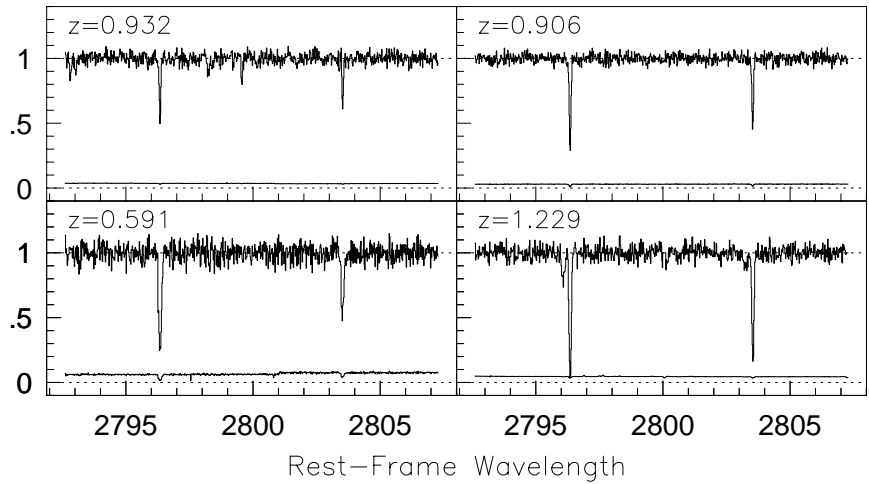,height=7.cm}
}}
\caption[]{Four weak Mg~II absorbing systems (presented in the rest
frame) at redshifts between 0.5 and 1.3.  The data were obtained with
the HIRES instrument at the W. M. Keck Observatory.}
\end{figure}

\section{Inferred Properties}

Details of the results presented here are given in Rigby, Charlton, \&
Churchill (2001).  In that work, we found two classes of SCWS: those
that are Fe--rich and those that are Fe--poor (see Figure~2),
demarcated by $\log N({\rm Fe~II})/N({\rm Mg~II}) = -0.3$.
We summarize our results as follows:

\begin{itemize}
\item The metallicities are greater than $0.1 Z_{\odot}$.
\item The Fe--rich systems have a low ionization parameter ($\log U =
\log n_{\gamma}/n_{\rm H}$) of $\simeq -4.5$, high density of $\log
n_{\rm H} \simeq -1$~cm$^{-3}$ (assuming $[\alpha/{\rm Fe}] = 0$), and
small cloud sizes of $d=10$~pc.
\item The Fe--poor clouds are less constrained and have $-4 < \log U
-2$ and $-3.5 < \log n_{\rm H} < 1.5$~cm$^{-3}$ (where can be in range
$[\alpha/{\rm Fe}] = 0$ to $+0.5$), and sizes in the range of $10 < d
< {\rm few~kpc}$.
\item Seven of the 15 systems require multiple ionization phases, due
to C~IV and/or Lyman~$\alpha$ absorption strengths.
\end{itemize}

\begin{figure}[h]
\centerline{\vbox{
\psfig{figure=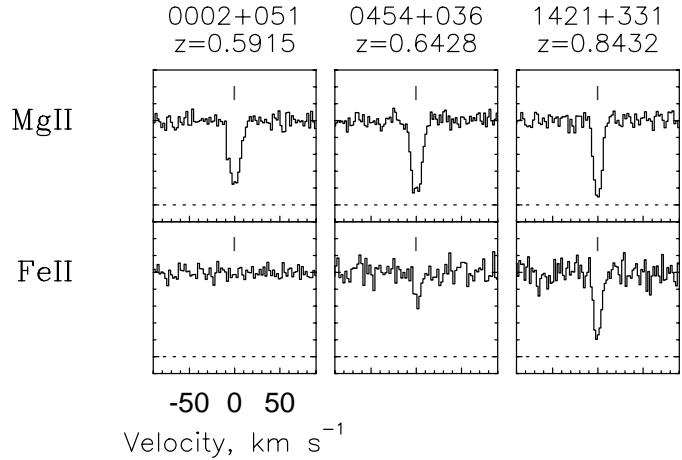,height=6.5cm}
}}
\caption[]{Example of three weak Mg~II absorbing systems, each with a
very different Fe II strength.  The $z=0.59$ system would be
considered Fe--poor and the other two Fe--rich.}
\end{figure}

\section{Astronomical Implications}

Because weak systems are optically thin to neutral hydrogen, we know
that their maximum allowed $\log N({\rm H~I})$ is 16.8~cm$^{-2}$.  If
we enforce the clouds to have no greater than solar metallicity, then
the cloud column densities cannot be much less than $\log N({\rm H~I}) =
15.8$~cm$^{-2}$.  Allowing for the uncertainty in the slope of the
neutral hydrogen column density distribution over this range of column
densities (Weymann et~al.\ 1998), we find that somewhere between
25--100\% of the $z\sim1$ Lyman~$\alpha$ forest is significantly metal
enriched (greater than 0.1 $Z_{\odot}$; also see Churchill \& Le~Brun 1998).

\begin{figure}[t]
\centerline{\vbox{
\psfig{figure=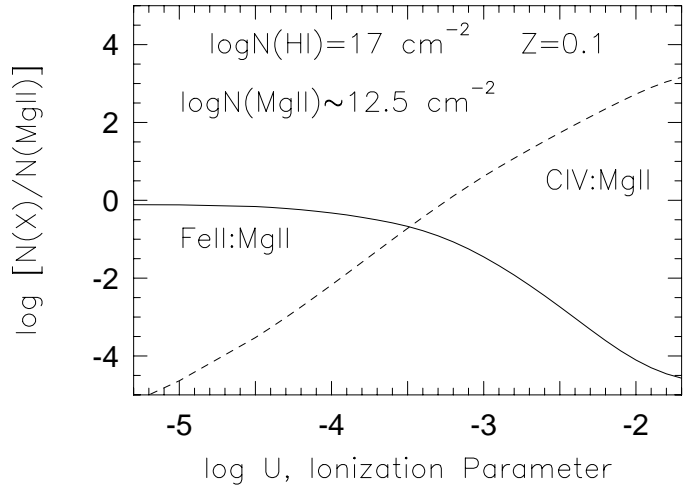,height=7.cm}
}}
\caption[]{Column density ratios with ionization
parameter for Fe~II and C~IV relative to Mg~II.  For a
near unity ratio of Fe~II to Mg~II, the ionization parameter is very
low,corresponding to a gas density of $\sim 0.1$~cm$^{-3}$.}
\end{figure}
\begin{equation}
\frac{n_{ws}}{n_{gal}} = {\rm few} \times 10^6 h^{-2}~~{\rm Mpc}^{-3}.
\end{equation}

If we consider the sub--set of systems which are Fe--rich, we find an
astronishing result about the spatial number densities, $n$.  Based upon
their redshift number densities, $dN/dz$, we find
\begin{equation}
n = 10^7 \left( \frac{\rm 1 pc}{R} \right)^{2} h~{\rm Mpc}^{-3} ,
\end{equation}
where $R$ is the cloud size.  The Fe--rich systems have very low
ionization conditions (or the Fe~II would not survive, see Figure~3).
This means that $N({\rm H~I}) \sim 0.01 N({\rm H})$ and that the
clouds have relatively high densities, $n_{\rm H} \sim 0.1$~cm$^{-3}$.
The cloud size scales as $R=N({\rm H})/n_{\rm H}$.  For
low--ionization, sub--Lyman limit systems, we have $R\simeq 10$~pc.
Thus, from Equation~1, we find $n \simeq 10^5 h~{\rm Mpc}^{-3}$.
Galaxies have $n \simeq 0.04 h^3~{\rm Mpc}^{-3}$. We infer that the 
ratio of Fe--rich weak systems to galaxies is

Because of the strong presence of iron and high metallicities, the
objects hosting the observed absorption very likely experienced
advanced stages of star formation (i.e.\ $\sim 1$ Gyr with
contributions from Type Ia SNe).  Are we tracking the elusive, old
Population III star clusters, or the theoretically predicted dark
mini-halos?

\vfill
\end{document}